\documentclass{webofc}

\usepackage[varg]{txfonts}   
\usepackage{hyperref}
\usepackage{url}

\usepackage{etoolbox}

\usepackage{amsmath}
\usepackage{listings}
\usepackage[toc,page]{appendix}
\usepackage{enumitem}
\setlist{nosep}

\hypersetup{colorlinks=true,citecolor=blue,urlcolor=blue,linkcolor=blue}
\begin{document}
\title{A lightweight analysis farm for fundamental physics experiments}

\author{\firstname{Sebastian} \lastname{Brommer}\protect\inst{1} \and \firstname{Ralf Florian} \lastname{von Cube}\inst{1} \and \firstname{Manuel} \lastname{Giffels}\inst{1} \and \firstname{Robin} \lastname{Hofsaess}\inst{1} \and \firstname{Markus} \lastname{Klute}\inst{1}\and \firstname{Benedikt} \lastname{Maier}\inst{2} \and \firstname{Raquel} \lastname{Quishpe}\inst{1} \and \firstname{Matthias} \lastname{Schnepf}\inst{1} \and \firstname{Luca} \lastname{Scotto Lavina}\inst{3} \and \firstname{Kathrin} \lastname{Valerius}\inst{1}}

\authorrunning{S.~Brommer \and M.~Giffels \and M.~Klute \and B.~Maier \and R.~Quishpe \and K.~Valerius}

\institute{Karlsruhe Institute of Technology, Karlsruhe, 76131, Germany
\and
           Imperial College, London, United Kingdom 
\and
           LPNHE, Sorbonne Universit\'e, Paris, 75252, France
          }

\abstract{

  Scientific collaborations require a strong computing infrastructure to successfully process and analyze data. While large-scale collaborations have access to resources such as Analysis Facilities, small-scale collaborations often lack the resources to establish and maintain such an infrastructure and instead operate with fragmented analysis environments, resulting in inefficiencies, hindering reproducibility and thus creating additional challenges for the collaboration that are not related to the experiment itself. We present a scalable, lightweight and maintainable Analysis Facility developed for the DARWIN collaboration as an example study case. Grid computing and storage resources are integrated into the facility, allowing for distributed computing and a common entry point for storage.
  The authentication and authorization infrastructure for all services is token-based, using an Indigo IAM instance.  
  We discuss the architecture of the facility, its provided services, the user experience, and how it can serve as a sustainable blueprint for small-scale collaborations.
}

\date{Received: date / Accepted: date}

%
\maketitle

\section{Introduction}

A robust computing infrastructure is essential for the success of scientific collaborations. Big computing sites like CERN or DESY offer feature-rich Analysis Facilities (AFs) to their user base, allowing large-scale collaborations such as the Large Hadron Collider (LHC) \cite{lhc} experiments to benefit from an infrastructure and services that provide integrated data, software and compute resources. Nevertheless, small-scale or new collaborations are not always affiliated to such computing sites, and may lack of resources to implement and maintain such infrastructure. Instead, they often rely on local university resources resulting in a more diverse computing infrastructure throughout the collaboration. This fragmentation may lead to inefficiencies, hinder reproducibility, and create collaboration challenges.

We propose a concept for a scalable, maintainable, lightweight AF, that can be deployed and operated with minimal expense.
It allows small-scale collaborations or university groups to run their own AF, while not relying on existing facilities. 
Using the DARk matter WImp search with liquid xenoN (DARWIN) project \cite{DARWIN_Aalbers_2016} as a study case, a prototype AF designed to be a blueprint for small-scale collaborations is presented. 

Since the deployment of the AF, the DARWIN project joined the XLZD collaboration.



\section{Requirements}
In general, it is difficult to formulate universally applicable requirements, as each research project and group has its own peculiarities.
The proposed AF is meant to be a prototype for an experiment-agnostic computing facility that can be utilized for all types of physics experiments, as it covers a large number of typical use cases, especially of young collaborations without a standardized computing concept.
It is designed to be a lightweight solution focusing on easy deployment and maintenance as well as good scalability to keep pace with increasing computing needs as new collaborations develop.

Particular attention is paid to ensuring that the concept presented meets the needs of all typical user groups.
To achieve this, the requirements from the user, as well as the administrative and operational point of view, are discussed in detail below.

\subsection{User requirements}
For such an AF, users can be categorized into two distinct groups: analysts and users responsible for central production.
Although their roles differ, their computing needs largely overlap.
While central production users require access to storage and compute resources to initially prepare, refine, and provide the experiment's data to all collaborators, the analysis users require in principle the same to develop and conduct their analyses.
The following list summarizes the common needs:

\begin{enumerate}
  \item \textbf{Single sign-on (SSO) solution}: Ensure that users can access all services within the AF using a single set of credentials, enhancing both security and user convenience.
  \item \textbf{Access methods}: Allow direct access via SSH for command-line interface and low-level system access but also offer interactive access through a web interface, allowing users to work with data and run analyses from any device with a web browser.
  \item \textbf{Local storage and compute resources}: Provide users with a way to do development work and to run small-scale tests without impacting shared resources.
  \item \textbf{Remote storage and compute resources}: Support scale-out of analysis work and central data processing tasks that may require significant computational power and remote storage.
  \item \textbf{Software environment management}: Offer tools and mechanisms for users to create, share, and manage software environments, supporting reproducible and portable software setups.
  \item \textbf{Monitoring and documentation}: Provide information about available resources, usage statistics, and detailed documentation on how to use various services of the AF.
\end{enumerate}

\subsection{Administrative requirements}

From the administration perspective, the AF should require minimal administrative overhead during operation. Consequently, the following requirements are identified: 

\begin{enumerate}
  \item \textbf{User management and authentication}: Automated account creation and deactivation processes, ideally integrated with the collaboration's existing user management systems\footnote{To ensure a clear separation of responsibilities for the user management and the technical operation of such an AF}.
  Automatic setup of (local) user accounts, home directories, and necessary access permissions according to predefined rules when a new user is added.
  \item \textbf{Transparent management of access rights and permissions}: Provide granular control over resource access, allowing easy assignment and modification of user privileges based on roles within the collaboration.
  \item \textbf{Scaling of direct-access resources}: Support an easy addition of local compute or storage nodes, with automated configuration and integration into the AF.
  \item \textbf{Dynamic integration of external compute resources}: Allow a dynamic incorporation of external, distributed resources, such as grid computing facilities or cloud providers, to expand computing capacity according to the demands.
  \item \textbf{Monitoring}: Provide insights into resource utilization, job statuses, and overall health of the AF for administrators.
  \item \textbf{Automated deployment}: Provide automated installation and configuration procedures of all necessary components to maintain reproducibility and to allow a quick setup of new instances of the AF.
\end{enumerate}

\section{Concepts and Implementation}
The implementation approach re-uses and builds upon concepts developed in a project for the PUNCH4NFDI Consortium \cite{punch4nfdi}.

\begin{figure*}
  \centering
  \includegraphics[width=0.999\textwidth]{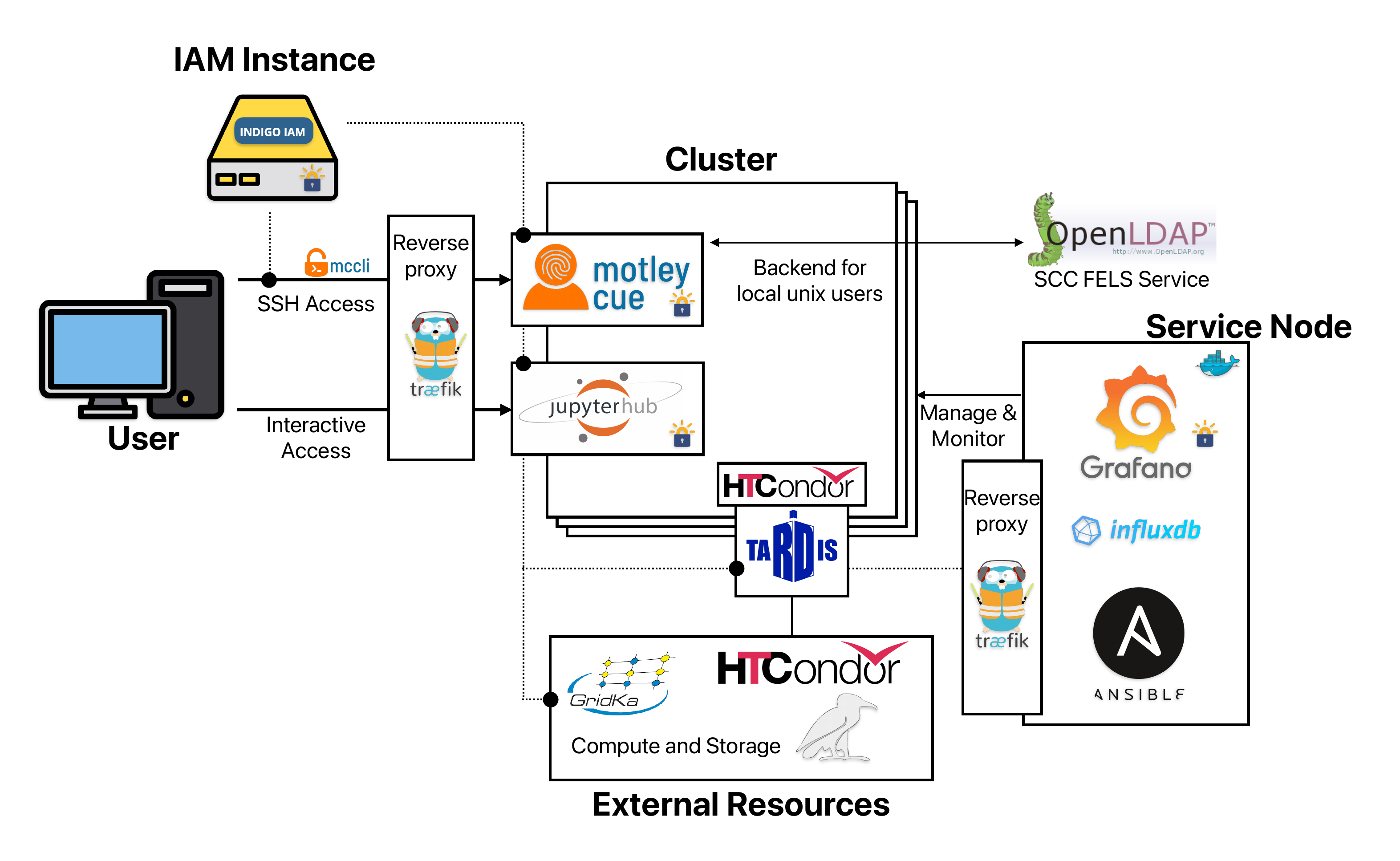}
  \caption{A schematic overview of the prototype AF. The prototype consists of a single cluster node that offers local storage and computational resources for development. Users can access the AF directly through SSH or interactively via JupyterHub. All authentication steps are managed by the IAM SSO. The figure’s dashed lines illustrate potential OAuth flows. Additional computational resources are integrated into an overlay batch system, which is based on HTCondor. A service node is employed to manage and monitor the entire AF.}
  \label{fig:facility}
\end{figure*}

\subsection{Hardware and Network Infrastructure}
The smallest possible setup of the AF consists of a single node that also provides direct-access storage, as well as a service node used for monitoring and deployment. The AF can be scaled by adding additional nodes to the cluster for direct-access resources and dedicated file servers for shared direct-access storage. The prototype AF described here is running with a minimal setup but will be extended to multiple nodes in the future. A technical overview of the prototype AF deployed is shown in Figure~\ref{fig:facility}.

\subsection{Authentication and Authorization Infrastructure}
The Authentication and Authorization Infrastructure (AAI) providing the Identity and Access Management (IAM) is realized using an IAM Indigo\cite{indigo_iam} instance. New users can register with the IAM instance via eduGAIN, which provides trusted proof of the identity for new users. After the registration, an administrator of the collaboration must verify that the newly registered user is a member of the collaboration and approve the new user registration. Alternatively, the collaboration membership information can come from a collaboration database that just requires an interface for the approval process.

All services of the AF can rely on the IAM to provide authentication and authorization of users via OpenID Connect (OIDC), which is built on top of OAuth2 \cite{oauth2}, using JSON Web Tokens (JWT) \cite{jwt}. Since the IAM provides all required API endpoints for OAuth2 and OIDC workflows, it is used as SSO provider for all services of the facility. Any changes in user permissions are propagated to all services of the facility whenever a user is required to authenticate.

Through the usage of groups within the IAM, specific access and permission rights of users can be managed. Due to the short-lived nature of the JWT, the facility can revoke the access rights of users promptly. Group memberships allow to split the user base into several distinct groups, like regular users, people responsible for central production and administrators.

\subsection{Facility access}
The AF provides direct SSH access, where users can authenticate using \textsc{JWT} via the \textsc{mccli} and \textsc{motley-cue} packages \cite{motley_cue,mccli}.

On the client side, the user must obtain a valid access token (e.g., using \textsc{oidc-agent}) and can then use this token with the SSH wrapper \textsc{mccli}. On the cluster side, the token is validated by a \textsc{motley-cue} server using the IAM API. If the user is attempting to connect to the cluster for the first time, a new UNIX account and all relevant directories are automatically created based on the information supplied by the IAM. All group memberships in the IAM are translated into UNIX user groups. The prototype AF uses an external LDAP server as a backend for UNIX user management, which allows the mapping of users to their UNIX account across all nodes within the facility.

For every subsequent login, the access token provided by the user is always linked to the same local account, and group memberships are updated according to the IAM information. If a user is entirely removed, they will no longer be able to log in. This procedure allows for transparent management of access rights and permissions, eliminating the need for manual user management on the facility side. No additional password is required.

\subsection{Compute resources and scalability}

For interactive analysis, a JupyterHub server is deployed and the authentication is handled via an \texttt{OAuth2} flow. Individual Jupyter notebooks of the users are run within Docker containers,  spawned by JupyterHub. Users can select from a set of predefined, validated and tested containers provided by the facility. Resource requirements, like CPU and memory, are specified by the user before launching the notebook. In the notebook, the user has access to their local storage. In addition, alternative approaches, such as using a VS Code server, are provided, allowing for a more IDE-like experience.

To provide computing resources to the user, an overlay batch system (OBS) using HTCondor is operated on the AF. The meta-scheduling solution COBalD/TARDIS \cite{2020EPJWC.24507040F,cobald,tardis} is used to allocate resources to external computing providers, that are integrated into the OBS using placeholder jobs. This practice is commonly used in the Worldwide LHC Computing Grid (WLCG) and allows to dynamically acquire computing resources based on demand while utilizing multiple resources. This creates a common entry point to compute resources for all AF users and a homogeneous computing infrastructure for the whole collaboration. Resources from the GridKa Tier-1 Computing Center are integrated into the prototype AF. A sketch of the scalability concept is shown in Figure \ref{fig:scale}.

\subsection{Storage solutions}

In addition to local storage for development work, grid storage can be utilized as the main data storage and to make the data available outside of the AF. The data can be accessed using storage protocols like \textsc{XRootD} \cite{xrd} or \textsc{WebDAV} \cite{webdav} with access tokens from the IAM instance for authentication. An additional proxy token service called \textsc{mytoken} \cite{mytoken,mytoken_server} together with an automatic renewal mechanism within HTCondor \cite{punch4nfdi} is used to ensure, that a valid access token is always available within long-running jobs. New access tokens are automatically generated on the submit node and then injected into the job, ensuring that the refresh tokens never leave the authenticated node. For the prototype described, a dCache hosted by the GridKa Tier-1 Computing Center is used.

\subsection{Software stack management}

The AF offers two primary, interconnected sources of software stack technologies. Firstly, it provides installations of \textsc{Apptainer} and \textsc{Docker}, which enables users to create and containerize their software configurations. Secondly, the read-only CernVM File System (\textsc{CVMFS}) \cite{cvmfs} is available to deliver read-only software stacks. As \textsc{CVMFS} is widely used in the grid, the identical software stack can be utilized both within batch jobs on the grid and the AF during development. Users can merge these two methods using tools like \textsc{cvmfs-unpacked} together with the \textsc{DUCC} service \cite{DUCC}, allowing them to distribute their containerized software stack to all necessary locations via \textsc{CVMFS}. For interactive analysis, these containerized images can be employed as well.

\subsection{Site Monitoring}
To continuously monitor and maintain a quick overview of the status of the facility, a monitoring stack based on \textsc{Grafana} and \textsc{InfluxDB} is used. Information on the status of the cluster nodes, storage, docker containers and the batch system is collected and visualized with a set of dashboards. Alert rules for critical components are defined, to notify the administrators in case of a failure.

\begin{figure*}
\centering
\includegraphics[width=0.975\textwidth]{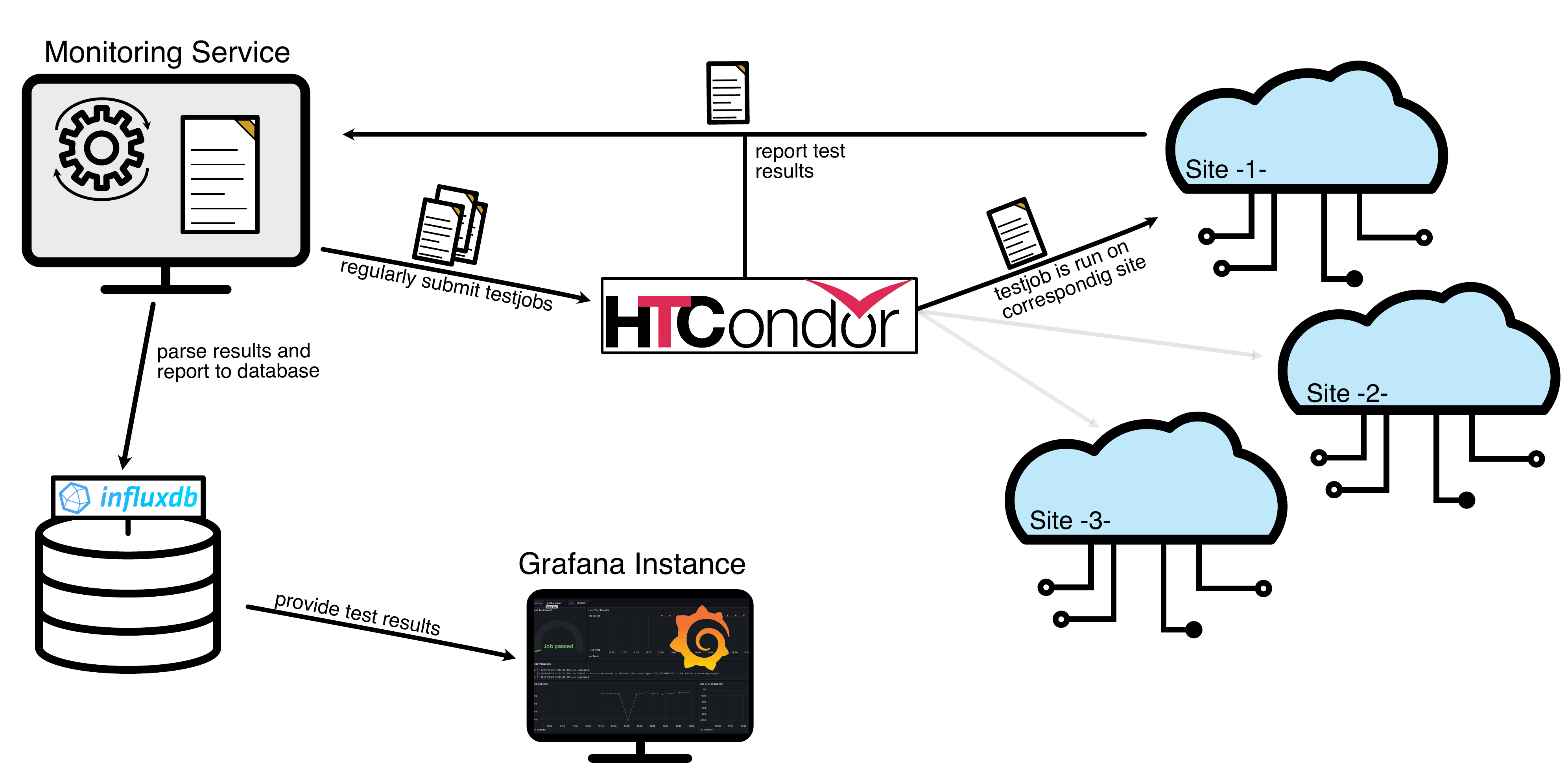}
\caption[Sketch of the Monitoring Service]{A sketch of the monitoring service. The service submits test jobs to individual sites, based on a given interval. Within the test, services like storage access or processing power can be tested. The results are then collected and visualized in a dashboard.}
\label{fig:monitoring}
\end{figure*}

The same monitoring stack is used, to monitor the availability and accessibility of compute resources connected to the facility. To automate this resource testing, a lightweight Python framework was developed, that is flexible enough to be used for any kind of grid resources and can test an arbitrary amount of different functionalities. A sketch of this monitoring service is shown in Figure \ref{fig:monitoring}.

\section{Experience and Future Plans}
The AF prototype has been made accessible to the DARWIN collaboration by December 2023, with initial utilization by a select group of users with the purpose of testing and feature enhancement.
Throughout this initial phase, the system has demonstrated stable performance without any unplanned outages, indicating its reliability and robustness.

\begin{figure*}
\centering
\includegraphics[width=0.999\textwidth]{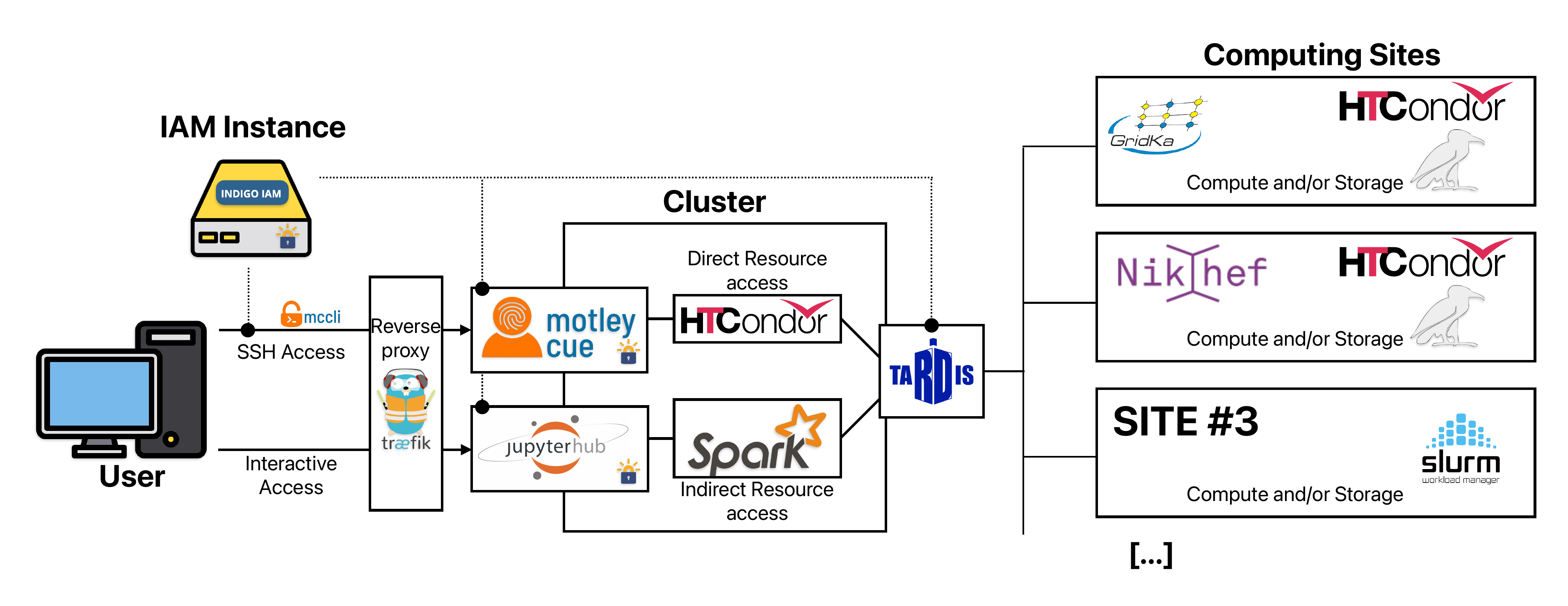}
\caption[Sketch of the scalability of the facility]{A sketch of the scalability of the facility. The computing sites shown are an example of what was implemented for the DARWIN collaboration study case. Additional compute resources are dynamically integrated using the meta-scheduler COBalD/TARDIS.}
\label{fig:scale}
\end{figure*}

Taking advantage of the scalability of the AF, additional computing sites will be integrated in the future in this study case, allowing the facility to provide enough computing resources for a large user base. In addition, it is foreseen to provide the resources for interactive analysis tasks via Jupiter also from external computing sites, following a similar approach to the SWAN (Service for Web based ANalysis)  service \cite{Piparo:2158559}. Interactive resources are managed in a Spark cluster, where the meta-scheduler COBalD/TARDIS is used to dynamically scale the size of the Spark cluster based on demand. A sketch of this foreseen infrastructure is shown in Figure \ref{fig:scale}.

\section{Summary}

A lightweight and scalable Analysis Facility (AF) designed for fundamental physics experiments has been presented. This AF addresses the common challenges faced by smaller scientific collaborations in establishing and maintaining a robust computing infrastructure. Therefore, a prototype of the AF was implemented using the DARWIN experiment collaboration as a study case.

The proposed AF is designed with minimal administrative overhead, making it easily deployable and maintainable, thus suitable for collaborations with limited IT resources. It provides a unified access point for all collaboration members, supporting both traditional computing methods and interactive analysis. The facility's scalable architecture integrates local resources with remote computing and storage resources, allowing for dynamic scaling based on demand. Security and user management are streamlined through a token-based authentication system utilizing an Indigo IAM instance, which implements a SSO solution across all services. 

The AF design presented in this work serves as a blueprint for other small-scale scientific collaborations, offering a sustainable solution to create a common analysis environment. As scientific collaborations continue to grow in size and complexity, solutions like this AF will become increasingly valuable in fostering effective collaboration and accelerating scientific discovery.\\




\begin{acknowledgement}    
We acknowledge the support of the Alexander von Humboldt Foundation, GridKa at the Karlsruhe Institute of Technology, Germany, and the KIT Center Elementary Particle and Astroparticle Physics (KCETA).
\end{acknowledgement}


\bibliography{biblio}

\end{document}